\documentclass[11pt]{article}

\usepackage{mathrsfs}

\evensidemargin=\oddsidemargin
\def\slash#1{\rlap/#1}
\def\tr{\mathop{\rm tr}}
\def\Eqn#1{Eq.~(\ref{#1})}
\def\Eqs#1#2{Eqs.~(\ref{#1}) and (\ref{#2})}

\title{\rightline{\normalsize DO-TH 11/07} \bigskip
\bf CPT-violating effects in muon decay}
\author{
{\bf Sebastian Hollenberg} \\ 
Fakult\"at f\"ur Physik, Technische Universit\"at Dortmund,\\ 
D-44221 Dortmund, Germany\\
\and \\
{\bf Palash B. Pal} \\
Saha Institute of Nuclear Physics, Kolkata, India}

\date{}

\begin{document}

\maketitle

\begin{abstract}
  We consider low-energy CPT-violating modifications in charged
  current weak interactions and analyze possible ramifications in muon
  and antimuon decays. We calculate the lifetime of muon and antimuon
  with these modifications, and from the result, put bounds on the
  CPT-violating parameters. Moreover, we elaborate on the muon and
  antimuon decay rate differentials in electron energy and spatial
  angle, which entail interesting phenomenological consequences
  presenting new ways to constrain CPT violation in charged lepton
  decays.
\end{abstract}

CPT invariance is one of the cornerstones of relativistic field
theory.  Under very general conditions outlined later, it is seen that
all Poincar\'e-invariant field theories are CPT-invariant.
Consequences of CPT invariance, like equality of mass and lifetime of
a particle and its antiparticle, have been tested to fairly good
accuracy~\cite{pdg2010}.

And yet, no matter how much theoretical prejudice goes in its favor,
the question of CPT invariance should be decided by experiments.
There might be unseen channels where effects of CPT violation might
show up at an observable level, or minute violations might manifest in
observables which have been measured, at a level below the present
limit of accuracy.  In fact, there is also some experimental evidence
that neutrino oscillation data favors different mixings in the
neutrino sector as opposed to the antineutrino
sector~\cite{AguilarArevalo:2010wv}.

In this paper, we consider low-energy CPT-violating effects in muon
and antimuon decays. We introduce CPT-violating vertex operators in
the standard model charged current interactions so that lifetimes of
particles and antiparticles are different. From experimental bounds on
the lifetimes, we put bounds on the CPT-violating couplings. Moreover,
we study decay rates differential in electron energy and spatial
angles and find that they also provide suitable new observables which
can further constrain CPT violation in charged lepton decays.

The assumptions which go into the proof of the CPT theorem are very
general, like the behavior of fields is goverened by a local
Lagrangian that is invariant under the proper Lorentz group, and the
fields with integer and half-integer spins obey Bose-Einstein and
Fermi-Dirac statistics respectively.  Violating any of these
conditions would lead to a complete reformulation of quantum field
theory, and would necessitate the introduction of new fields with new
associated particles~\cite{Colladay:1996iz,Kostelecky:2000mm}.

We take a more conservative approach to CPT violation.  To appreciate
our viewpoint, let us give a quick and easy review of the proof of the
CPT theorem \cite{Lahiri:2005sm}.  A vector field like the photon is
odd under CPT, as can be easily seen for the photon field which is odd
under each of the operations C, P, T.  All scalar fields can be defined
to be even under CPT transformations.  Fermion fields must appear
in bilinear combinations in a Lagrangian, so it is enough to consider
the CPT properties of such bilinears.  It can be seen that the
bilinears involving an odd number of Dirac matrices are odd under CPT,
whereas those with an even number of Dirac matrices are even under
CPT.  Thus, for scalars, fermions as well as vector fields, we can
make the general statement that the field operators (or their
combinations) with odd (even) number of Lorentz vector indices are
respectively odd (even) under CPT.  To make the discussion transparent,
let us assume that all of these indices are contravariant indices.  In
a Lagrangian, these indices have to be contracted by some tensors
inherent of spacetime.  In the 4-dimensional Minkowski spacetime, the
only properties of spacetime that can help in the contraction of
indices are the metric tensor $g_{\alpha\beta}$ and the completely
antisymmetric tensor $\varepsilon_{\alpha\beta\lambda\rho}$.  Thus,
the number of indices carried by fields and bilinears must be even,
and therefore the Lagrangian must be even under CPT.

The argument does not hold if spacetime is endowed with some
characteristic tensors of odd rank. The effects on physics due to the
presence of such tensors in the Dirac equation has been discussed by
Kosteleck\'y and collaborators \cite{Kostelecky:2003cr,
  Kostelecky:2003xn}.  They showed, among other things, that
CPT-violating effects can arise from terms involving such objects, and
discussed how these effects manifest, e.g., in the masses and
oscillations of neutrinos.

The paradigm of our analysis here is a different one.  We assume that
the free Dirac equation is not altered by the presence of CPT
violation; only some interactions violate CPT.

Since the upcoming analysis deals with muon decays
altered by CPT-violating interactions, we are interested in
the charged current part of the standard model, which is given by
\begin{eqnarray}
\mathscr L_{\rm cc} = - {g \over \surd2} J^\lambda W_\lambda +
\mbox{h.c.} 
%\label{}
\end{eqnarray}
In the standard model, the current in the leptonic sector is given by
\begin{eqnarray}
J^\lambda = \sum_{\ell=e,\mu,\tau} \; \bar\ell \gamma^\lambda L \nu_\ell \,,
%\label{}
\end{eqnarray}
where
\begin{eqnarray}
L \equiv \frac12 (1-\gamma_5)
%\label{}
\end{eqnarray}
is a chirality projection operator.  We will entertain the idea that
the expression for $J^\lambda$ is enhanced by additional tensors of
odd rank in order to establish CPT violation in the
interaction. Hence, we substitute
\begin{eqnarray}
J^\lambda \to J^\lambda \equiv \sum_{\ell=e,\mu,\tau} \; 
\bar\ell \Big( \gamma^\lambda L + \delta \Gamma^\lambda
\Big) \nu_\ell 
%\equiv \sum_{\ell=e,\mu,\tau} \; 
%\bar\ell \Gamma^\lambda \nu_\ell \,,
%\label{}
\end{eqnarray}
and consider the possibility that 
some of these {\em extra terms} $\delta\Gamma^\lambda$ are CPT-violating.  
Because of experimental constraints that exist, e.g. lifetime of the muon and
antimuon, the effect of these extra terms has to be very small. We
shall see in due course how to explicitly parametrize the additional
coupling $\delta\Gamma^\lambda$ and actually quantify the smallness of
the odd-rank tensors to be introduced. The fact that experimental
constraints exist, leads us to consider only the first order effects
of $\delta\Gamma^\lambda$; we neglect all contributions
``$\mathcal{O}(\delta\Gamma^2)$''.

Having laid down our paradigm for
CPT violation in leptonic currents, we can now delve into applications
and explore its consequences on observables such as muon
and antimuon lifetimes.
We begin
by assigning momenta for the particles involved in muon decay as follows:
\begin{eqnarray}
\mu^-(p) \to e^-(p') + \nu_{\mu}(k) + \bar{\nu}_{e}(k') \,.
%\label{}
\end{eqnarray}
For the antimuon decay, the notation for the momenta will be the same
for the corresponding antiparticles. Since all masses are much
smaller compared to the $W$-boson mass, we can write the Feynman
amplitude of the muon decay process as
\begin{eqnarray}
\mathscr M (\mu^- \to e^- \nu_{\mu} \bar{\nu}_e) 
   = 2\surd2 G_{\rm F} 
     \Big[ \bar{u}(p')\Gamma^\lambda v(k') \Big] 
     \Big[ \bar{u}(k)\Gamma_\lambda u(p) \Big],
\label{M}
\end{eqnarray}
using $\Gamma^\lambda \equiv \gamma^\lambda L + \delta \Gamma^\lambda$
as a shorthand notation in order to streamline notation.  
The squared spin-averaged matrix element for
the muon decay can now be written as
\begin{eqnarray}
\left< {\vert\mathscr M(\mu^- \to e^- \bar{\nu}_e \nu_{\mu})\vert^2}
\right> 
&\equiv& \frac12 \sum_{\rm spins}
    \vert\mathscr M(\mu^- \to e^- \bar{\nu}_e \nu_{\mu})\vert^2 
\nonumber\\* 
&=& 4G_{\rm F}^2 \tr \Big\{\Gamma^\lambda (\slash p + m_\mu) 
\overline\Gamma^\rho \slash k \Big\} 
     \tr \Big\{\Gamma_\lambda \slash k'
     \overline\Gamma_\rho \slash p' \Big\} \,,
%\label{}
\end{eqnarray}
where $\overline\Gamma^\lambda = \gamma_0 {\Gamma^\lambda}^\dagger
\gamma_0$ is the Dirac adjoint, and $m_\mu$ is the muon mass.  We neglect the masses of all
decay particles in what follows.

For the $\mu^+$-decay, the Feynman amplitude can be obtained by
replacing any $u$-spinor by the corresponding $v$-spinor and vice
versa.  The only difference in the value of $\left< \vert{\mathscr M\vert^2} \right> $ 
would be that the sign of the mass term would be reversed,
since it would come from the spin sum of $v$-spinors of the antimuon.
This observation suggests that CPT-violating effects can be best {\em
  isolated} from the standard model interaction by taking the
difference in decay rates for muon and antimuon. Put another way, CPT
violation manifests in the different lifetimes for the muon and
antimuon. Following this train of thought we introduce the difference
in spin-averaged matrix elements squared $\delta \mathscr M^2$ for
muon and antimuon decays.  We obtain
\begin{eqnarray}
\delta \mathscr M^2 &\equiv& 
\left< {\vert\mathscr M(\mu^- \to e^- \bar{\nu}_e \nu_{\mu})\vert^2}
\right> 
- \left< {\vert\mathscr M(\mu^+ \to e^+ \nu_e
  \bar{\nu}_{\mu})\vert^2} \right> 
\nonumber\\
&=& 8 G_{\rm F}^2 m_\mu  \tr \Big\{
\Gamma^\lambda \overline\Gamma^\rho \slash k \Big\}
     \tr \Big\{\Gamma_\lambda \slash k'
     \overline\Gamma_\rho \slash p' \Big\} \,.
%\label{}
\end{eqnarray}
Clearly, this vanishes if $\delta\Gamma^\lambda=0$, because the first
trace then contains an odd number of Dirac matrices.

The situation changes if the current contains terms which have an even
number of Dirac matrices.  In that case, the interference of those
even terms with the usual $\rm (V-A)$ structure can give a non-zero value
for the trace in question.  Hence we take
\begin{eqnarray}
\delta \Gamma^\lambda = A^\lambda + {B^{\lambda}}_{\alpha\beta}
\sigma^{\alpha\beta} \,, 
%\label{}
\end{eqnarray}
where $A^\lambda$ and ${B^{\lambda}}_{\alpha\beta}$ are a set of real 
constants, parametrizing CPT violation.  Obviously, by definition,
${B^{\lambda}}_{\alpha\beta} = - {B^{\lambda}}_{\beta\alpha}$.
Henceforth, we will refer to $A^{\lambda}$ as the {\it vector part}
and $B^{\lambda}_{\,\,\,\alpha\beta}$ the {\it dipole part} of the
CPT-violating contributions.

To the first order in $\delta\Gamma^\lambda$, we can write
\begin{eqnarray}
\delta \mathscr M^2 &=& 16 G_{\rm F}^2 m_\mu \Big( k'_\lambda p'_\rho +
k'_\rho p'_\lambda - k' \cdot p'\, g_{\lambda\rho} - i
\varepsilon_{\lambda\alpha\rho\beta} k'^\alpha p'^\beta \Big)
\nonumber\\* 
&& \times \tr \Big\{ \delta\Gamma^\rho 
\gamma^\lambda L \slash k + \gamma^\rho L 
\delta\Gamma^\lambda \slash k \Big\} \,,
\label{dMsq}
\end{eqnarray}
where the Levi-Civita tensor has been defined with the convention
$\varepsilon_{0123} = +1$.  The calculation of the remaining trace is
easy, and the results obviously contain just a single power of the
momentum $k$.

Phase space integration then yields the difference in decay rates
$\Delta\Gamma$ for muons and antimuons
\begin{eqnarray}
   \Delta\Gamma = \frac{G^2_{\mathrm F}}{2\pi^5} 
\int\frac{\mathrm d^3 p'}{2E'}  I_{\alpha\beta}(q)
\left[T_A^{\alpha\beta}(p') + T_B^{\alpha\beta}(p') \right],
\label{DGam}
\end{eqnarray}
where the CPT-violating contributions are absorbed into the
tensors
\begin{eqnarray}
    T_A^{\alpha\beta}(p') &=& p' \cdot A\, g^{\alpha\beta} \,, 
\label{TA}\\
   T_B^{\alpha\beta}(p') &=& \epsilon_{\lambda\rho\mu\nu} \left(2 B^{\lambda\alpha\rho} 
   g^{\beta\mu} p'^{\nu} +  (B^{\beta\mu\nu} p'^{\rho} -
   B^{\rho\mu\nu} p'^{\beta}) g^{\lambda \alpha}\right) \,, 
\label{TB}
\end{eqnarray}
and integration over the momenta of the two neutrinos is of the form
\begin{eqnarray}
I_{\alpha\beta} (q) = \int {\mathrm d^3k \over 2k_0} \int {\mathrm
  d^3k' \over 2k'_0} \; \delta^4(q-k-k')  k_\alpha k'_\beta \,, 
\label{I}
\end{eqnarray}
where $q=p-p'$. 
The neutrino phase space
integrals appear exactly in the form given in \Eqn{I} when one
calculates the muon decay rate in the standard model.  In view of the
fact that the expression in \Eqn{dMsq} is already linear in the
CPT-violating parameters, we can use the usual form
\cite{Lahiri:2005sm} of the integral:
\begin{eqnarray}
I_{\alpha\beta} (q) = {\pi \over 24} (q^2 g_{\alpha\beta} + 2q_\alpha
q_\beta) \,.
%\label{}
\end{eqnarray}
Note that $I_{\alpha\beta}$ is symmetric in its indices.  We
have used this property to eliminate the antisymmetric parts of the
tensors that appear in \Eqs{TA}{TB}.

The rest of the calculation is straightforward, and yields the
following expression for the difference in decay rates $\Delta\Gamma$
for muon and antimuon in their rest frame:
\begin{eqnarray}
\Delta\Gamma = {G_{\mathrm F}^2 m^5_\mu \over 192 \pi^3} \Big( A_0 -
\varepsilon_{0ijk} B^{ijk} \Big) \,.
%\label{}
\end{eqnarray}
From this expression it is readily seen that both the vector and
dipole parts violate CPT invariance and would hence contribute to the
difference in muon and antimuon lifetimes.  
It is also interesting to note that, although by definition the tensor $B$ is
antisymmetric in its last two indices only, it is the completely
antisymmetric part of the tensor that contributes to the decay rate.

Clearly, we see that the presence of odd-rank tensors inherent in
spacetime produces CPT-violating effects.  The magnitude of the
parameters can be restricted from the known bounds on lifetime
differences of the muon and the antimuon.  Using
\begin{eqnarray}
\frac{\tau(\mu^+)}{\tau(\mu^-)} = 1.00002 \pm 0.00008
%\label{}
\end{eqnarray}
to the 1$\sigma$ level~\cite{pdg2010}, we can set the bounds on the
CPT-violating parameters that we have used:
\begin{eqnarray}
A_0 < 10^{-4} \,, \qquad \varepsilon_{0ijk}
B^{ijk} < 10^{-4} \,.
%\label{}
\end{eqnarray}
Similar bounds can be obtained from tau lifetimes, but they are
somewhat less restrictive.

More information on CPT-violating parameters can be obtained if we
find the differential decay rate with respect to the energy of the
charged particle in the final state.  For this, we go back to
\Eqn{DGam} and integrate that equation with respect to the angular
variables.  For the vector part, this yields
\begin{eqnarray}
   \frac{\mathrm d \Delta\Gamma_{\mathrm A}}{\mathrm d x} =
   \frac{G^2_{\mathrm F}m^5_\mu}{16 \pi^3} 
   x^2 \left(1-x\right)  A_0.
\end{eqnarray}
Here $x$ is a dimensionless energy variable, defined by 
\begin{eqnarray}
x = \frac{2E'}{m_\mu} \,.
%\label{}
\end{eqnarray}
The distribution vanishes at the kinematic boundaries of $x=0$ and $x
= 1$. It attains a maximal value at $x_{\rm peak} = \frac{2}{3}$. Both
these properties are independent of the explicit CPT-violating
parameter $A_0$ and yet for $A_0 = 0$, i.e. in the absence of CPT
violation, the energy dependence of the difference in muon and
antimuon decay rates does not exist.  Put another way, CPT-violating
effects (here: a preferred direction) also shift the energy spectra of
electrons and positrons emergent from muon and antimuon decays
relative to one another. This difference is proportional to the time
component of the preferred 4-vector of spacetime.
Irrespective of the value of $A_0$, the
difference in spectra peaks at $x_{\rm peak}= \frac{2}{3}$ or
equivalently $E'_{\rm peak} = \frac{m_\mu}{3}$ provided the only
CPT-violating effects are coming from $A^\lambda$.

Now we include the contribution from the dipole part. 
We obtain
\begin{eqnarray}
   \frac{\mathrm d \Delta\Gamma_{\mathrm B}}{\mathrm d x} = - \frac{G^2_{\mathrm F}m^5_{\mu}}{48\pi^3} 
   x^2 \left(1-\frac{1}{3}x\right) \varepsilon_{0ijk} B^{ijk}.
\end{eqnarray}
This contribution to the
difference in energy distributions vanishes at $x=0$, but neither does it
vanish anywhere else, nor does peak within the kinematic region.

Summing both contributions stemming from the vector and the dipole
part we infer the following: the difference in electron and
positron energy spectra from muon and antimuon decay definitely
vanishes at $x=0$.  It may also vanish at $x = \frac{9A_0-3\eta}{9A_0-\eta}$
where $\eta=\varepsilon_{0ijk}B^{ijk}$ provided this value of $x$ is
within the kinematic region $0<x<1$.  The difference will be largest
at $x_{\rm peak} = \frac{6A_0-\eta}{9A_0-\eta}$ if this is within the
kinematic region; otherwise, it will be largest for $x=1$.

It should be noticed that the total decay rate cannot restrict in any way
the spatial components of $A^\lambda$, and the components of
$B^{\lambda\alpha\beta}$ with any of the indices equal to the time
component. 

However, we now show that
restrictions on these components of $A$ and $B$ which are not present
in the total decay rate can be obtained from considerations of the
decay rate differentials in the spatial angle $\mathrm d \Omega$.
To this end, we go back to \Eqn{DGam} and integrate over the magnitude
of the momentum $p'$.  For the vector part, it gives
\begin{eqnarray}
\frac{\mathrm d \Delta\Gamma_{\mathrm A}}{\mathrm d \Omega} 
= \frac{G^2_{\mathrm F}m_\mu^5}{768 \pi^4} 
\left(A_0 - |\vec A| \cos\vartheta\right) \,,
   \label{GAtheta}
\end{eqnarray}
where $\vartheta$ is the angle between the
electrons (positrons) emergent from the muon (antimuon) decays and the preferred direction $\vec{A}$.  Put
another way, not only does CPT violation enforce a slight difference
in energy spectra for electrons and positrons, but it also alters
their angular distributions with respect to one another.  The angular
dependence is proportional to the spatial components of $A^\lambda$.
The direction and magnitude of $\vec A$ can then in principle
be determined from the angular dependence given in \Eqn{GAtheta}.

The angular dependence for the dipole part is found to  
intricately depend on both the azimuth as well as the zenith angle:
\begin{eqnarray}
   \frac{\mathrm d \Delta\Gamma_{\mathrm B}}{\mathrm d \Omega} = -\frac{G^2_{\mathrm F}m^5_{\mu}}{192\pi^4}
   \left[\frac{5}{24} \varepsilon_{0ijk}B^{ijk} + \frac{5}{24} \varepsilon_{0ijk}B^{0jk} \hat{p}'^{i}
   - \frac{1}{8} \varepsilon_{i\kappa\lambda \rho} B^{\kappa\lambda\rho} \hat{p}'^{i}
   - \frac{1}{8} \varepsilon_{0ijk} B^{ljk} \hat{p}'_{l} \hat{p}'^{i}\right],
\end{eqnarray}
where $\hat{p}'^{i}$ is a unit vector which can be written in
spherical coordinates according to $\hat{p}'^{i} = (\sin\vartheta
\cos\phi, \sin\vartheta \sin\phi, \cos\vartheta)$. Two observations
are readily made: the dipole part shows a rich angular dependence;
statements about the time components of $B$ now become possible by analyzing
the decay rate differential in the spatial angles.

Both the vector and the dipole part reveal interesting
phenomenological consequences on their own. If both effects are to be
considered simultaneously, one again simply adds the
respective contributions.

Of course if CPT is violated, it can manifest in the muon and
antimuon decay in many possible ways~\cite{Bluhm:1999dx}.  
The masses of muon and antimuon
might be different, which would result in different decay rates.  In
this paper, we assumed that the free part of the Lagrangian is CPT
invariant, and CPT violation occurs only through interactions.  With
this scenario, we have extended the standard model charged current
weak interactions to include CPT-violating parameters.  We see, from
our analysis, that in principle the parameters can be determined by
measuring the total as well as differential rates of the decay of muon
and antimuon.

\paragraph*{Acknowledgements:} PBP wants to thank DAAD-DST PPP Grant
No. D/08/04933, and DST-DAAD Project No. INT/DAAD/P-181/2008 which
enabled him to visit Dortmund.  The question of CPT violation was
discussed during this trip between the authors, Heinrich P\"as and
Octavian Micu.  PBP wants to thank Anindya Datta for further
discussions; SH wants to acknowledge further discussion with Octavian
Micu and Danny van Dyk.


\begin{thebibliography}{99}\small\itemsep=0pt

\bibitem{pdg2010} K. Nakamura et al. (Particle Data Group), JPG 37,
  075021 (2010) (URL: http://pdg.lbl.gov)

%\cite{AguilarArevalo:2010wv}
\bibitem{AguilarArevalo:2010wv}
  A.~A.~Aguilar-Arevalo {\it et al.}  [The MiniBooNE Collaboration],
  %``Event Excess in the MiniBooNE Search for $\bar \nu_\mu \rightarrow \bar
  %\nu_e$ Oscillations,''
  Phys.\ Rev.\ Lett.\  {\bf 105} (2010) 181801
  [arXiv:1007.1150 [hep-ex]].
  %%CITATION = PRLTA,105,181801;%%

%\cite{Colladay:1996iz}
\bibitem{Colladay:1996iz}
  D.~Colladay, V.~A.~Kosteleck\'y,
  %``CPT violation and the standard model,''
  Phys.\ Rev.\  {\bf D55}, 6760-6774 (1997).
  [hep-ph/9703464].

%\cite{Kostelecky:2000mm}
\bibitem{Kostelecky:2000mm}
  V.~A.~Kosteleck\'y, R.~Lehnert,
  %``Stability, causality, and Lorentz and CPT violation,''
  Phys.\ Rev.\  {\bf D63}, 065008 (2001).
  [hep-th/0012060].

%\cite{Lahiri:2005sm}
\bibitem{Lahiri:2005sm} See, e.g., 
  A.~Lahiri, P.~B.~Pal,
  ``A first book of quantum field theory,''
  Narosa Publishing and Alpha Sci. Int. (2nd edition, 2005), Ch.~10.

%\cite{Kostelecky:2003cr}
\bibitem{Kostelecky:2003cr}
  V.~A.~Kosteleck\'y and M.~Mewes,
  %``Lorentz and CPT violation in neutrinos,''
  Phys.\ Rev.\  D {\bf 69} (2004) 016005
  [arXiv:hep-ph/0309025].
  %%CITATION = PHRVA,D69,016005;%%

%\cite{Kostelecky:2003xn}
\bibitem{Kostelecky:2003xn}
  V.~A.~Kosteleck\'y and M.~Mewes,
  %``Lorentz and CPT violation in the neutrino sector,''
  Phys.\ Rev.\  D {\bf 70} (2004) 031902
  [arXiv:hep-ph/0308300].
  %%CITATION = PHRVA,D70,031902;%%

%\cite{Bluhm:1999dx}
\bibitem{Bluhm:1999dx}
  R.~Bluhm, V.~A.~Kosteleck\'y, C.~D.~Lane,
  %``CPT and Lorentz tests with muons,''
  Phys.\ Rev.\ Lett.\  {\bf 84}, 1098-1101 (2000).
  [hep-ph/9912451].



\end{thebibliography}
\end{document}